# DESIGNING MOBILE HEALTH FOR USER ENGAGEMENT: THE IMPORTANCE OF SOCIO-TECHNICAL APPROACH

Tochukwu Ikwunne, ADAPT Centre, Trinity College Dublin, Ireland, ikwunnet@tcd.ie

Lucy Hederman, ADAPT Centre, Trinity College Dublin, Ireland, hederman@tcd.ie

P. J. Wall, ADAPT Centre, Trinity College Dublin, Ireland, pj.wall@tcd.ie

**Abstract:** Despite the significance of user engagement for efficacy of mobile health (mHealth) in the Global South, many such interventions do not include user-engaging attributes. This is because socio-technical aspects are frequently not considered during the design, development, and implementation, stages of such initiatives. In addition, there is little discussion in the literature about the role socio-technical factors play in user-centered design processes for mHealth. This research posits consideration of socio-technical factors is required as techno-centric approaches to mHealth design and user engagement, as well as those relying on existing universal frameworks for user-centered design, have proven to be ineffective with the result that most mHealth projects in the Global South fail to sustain. This research examines projects in Sierra Leone where semi-structured interviews were conducted with mHealth designers and developers in order to explore their attitudes towards user engagement in this case. Barriers and facilitators to user engagement were identified and classified as either technical or socio-technical. Findings from the study indicate that adoption of a techno-centric approach without consideration of socio-technical factors can negatively affect user's engagement. Based on these findings, we propose to develop a new design framework for more effective inclusion of user-engaging attributes in mHealth.

**Keywords:** mHealth, user engagement, socio-technical, techno-centric, user-centered design

## 1. INTRODUCTION

It is broadly accepted that the ubiquity of mobile phones in the Global South presents an opportunity to address many of the challenges posed by severely under-resourced public health and education systems. However, the design, development, implementation and use of systems leveraging mobile phones for healthcare (mHealth) has proven to be problematic for a variety of reasons, and a majority of such systems fail to sustain and scale (Heeks, 2016). There are many reasons put forward for this (Wall, 2013), but the lack of attention to user-engaging attributes has not received attention in the literature to date. This paper seeks to acknowledge that socio-technical factors responsible for user engagement in mHealth need to be considered as part of the design process. Socio-technical factors are used to represent broadly socio-cultural, and technical factors. According to (Whetton and Georgiou, 2010), a socio-technical perspective seeks to explore the organisational, professional, cultural and technical factors inherent in the design of information systems. A socio-technical approach includes consideration of both social and technical factors based upon combination of the human and the technology (Hadi et al., 2019). Moreover, according to Whetton (2005), a socio-technical approach seeks to identify the synergy between technology and the social, professional, and cultural environment in which it is used. This is because techno-centric approaches, and other approaches which rely on existing universal frameworks for user-centered design, have been proven to be ineffective (Tabi et al., 2019). For example, the literature discusses designers' assumptions during the design process (Cornet et al., 2020) resulting in a lack of user engagement with mHealth systems (Gopalakrishnan et al., 2020; Nicholas et al., 2017), as well as lack of socio-technical understanding of mHealth design (Aryana et., 2019). Based on this lack of attention to user-engaging attributes in mHealth, and specifically the absence of a focus on socio-technical factors in user engagement, we identified an important research gap. In an attempt to address this research gap, our research proposes that socio-technical factors responsible for user engagement need to be taken into account in the mHealth design and development process. We thus propose the following research question: how does consideration of socio-technical factors in the mHealth design and development process improve user engagement? In order to answer the research question, we examine an mHealth implementation in the Bonthe District of Sierra Leone. Data was collected by semi-structured interviews with mHealth designers and developers to explore their experiences of why users either engage or not with the mHealth application in this case. The objective is to identify facilitators and barriers to engagement with mHealth interventions through the lens of the COM-B model, and to classify these facilitators and barriers as either technical or socio-technical factors of user engagement. The outcome of the interviews to date indicates that the inclusion of socio-technical factors during mHealth design and development is a requirement for more effective user engagement with the technology.

The remainder of this paper proceeds as follows: Section 2 presents the literature review, with the mHealth case in Sierra Leone presented in Section 3. Section 4 details the methodology adopted for this research, and Findings are presented Section 5. Section 6 presents DECENT, the proposed mHealth design framework resulting from this work. The paper concludes with brief conclusions from the work to date in Section 7.

## 2. LITERATURE REVIEW

This section commences with an examination of the literature on user engagement before moving to examine the body of work relating to mHealth failure and underperformance in the Global South. In particular, attention is paid to the relevance of socio-technical factors in the design process of mHealth.

### 2.1 User Engagement

There is an extensive body of literature on the theme of user engagement with technology. The term "user engagement" is an all-inclusive phrase encompassing the relationship between users and the technology they are using. There are a variety of descriptions and definitions of the term provided, including by Attfield et al, (2011, p. 10) who defines user engagement as *"the emotional, cognitive and behavioral connection that exists, at any point in time and possibly over time, between a user and a resource"*. User engagement is important to mHealth, with many researchers (e.g. Schnall et al., 2016; Curtis et., 2015) proposing that the design process of mHealth should meet various users' needs. However, many current mHealth interventions are designed on existing healthcare system constructs (Verhoeven et al., 2010) which encourages designers to base their designs on assumptions that are not validated with primary user input (McCurdie et al., 2012). Thus, the resulting interventions are not as effective as those that involve end-users' needs (Verhoeven et al., 2010) and inputs from relevant stakeholders such as commercial app industries and experts in the design process (Curtis et al., 2015). Furthermore, McCurdie et al. (2012) state that user-centered design involves an approach which is informed by the needs and understanding of a specific end-user group and plays a key role in achieving user engagement with technology.

In addition to this, there is an extensive body of work discussing socio-technical and socio-cultural factors associated with user engagement (McCurdie et al. 2012), as well as the environment in which the intervention will be used. As stated in the Introduction, socio-technical factors are used to represent broadly socio-cultural, and technical factors. According to (Whetton and Georgiou, 2010), a socio-technical perspective seeks to explore the organisational, professional, cultural and technical factors inherent in the design of health care information systems. Wickramasinghe, (2018) posits that lack of engagement with mHealth systems is due to socio-cultural and organizational issues, for example, when mHealth applications developed in the Global North are implemented in the Global South where there may be many and varied social, cultural and political differences and beliefs. In cases such as this it is essential that implementation be localized to enhance success of the mHealth systems. This point is also emphasized by Shozi et al. (2012, p.1) who suggest that the assumption that technology designed in the Global North can be simply dropped into the Global South and expected to work is "*fallacy*". Furthermore, according to Manda, T. D., & Msosa (2011, p. 210), "*mHealth comprises multiple socio-technical arrangements, which, among others, include workers' information needs, workflow and usability requirements, available technology options, and how best technology can be adapted to suit these needs and requirements*". Thus, there is a need for a better understanding of the complexity of user needs and how to incorporate this information effectively into the design process of mHealth.

### 2.2 mHealth Failure and Underperformance in the Global South

There have been many reasons presented for the high level of mHealth failure and underperformance in the Global South. Writing many years ago, Kaasbøll & Nhampossa (2002) identified that socio-cultural considerations have a huge influence on the outcome of any health information system (HIS) implementation. They discussed the socio-cultural issues associated with transfer of HIS between the public health sectors of Mozambique and South Africa. The paper showed that socio-cultural differences between the two countries necessitated considerable adjustment and adaptation to cope with local variations. Similarly, a wide range of papers (e.g., Wu et al., 2007; Huang et al., 2019; Tate et al., 2013; Bentley et al., 2019; Ikwunne et al., 2020; Jacob et al., 2020; Woodland et al., 2021; Hofstede, 2011) have argued that successful implementations require better understanding of socio-technical practices of user groups and environmental differences because of their significance and impact. Additionally, it is recognized that a major reason for failure is unsuitable design as related to the needs and context of use (Braa, 2007). Koskinen (2017), posited that global technology carries meanings and structures that may or not fit with local realities and highlights the need for a framework for understanding context which contributes to the understanding of local technology production in under-resourced and developing contexts. Moreover, Wall et al. (2013) observed that the adoption of a techno-centric approach without consideration of socio-technical issues can negatively affect an mHealth implementation. A further reason for mHealth failure is the use of a top-down approach by implementers (Braun et al. 2013). Such an approach adopts a techno-centric style without allowing users to provide feedback to the technology that they expect to use. Despite this abundance of literature on mHealth and HIS failure and underperformance however, there is a little discussion concerning the role played by socio-technical factors in user-centered design processes for mHealth (Aryana et al., 2019; Farao et al., 2020).

## 3.      MOBILE HEALTH IN SIERRA LEONE

The research presented in this paper is based on an mHealth case study in the Bonthe District of Sierra Leone, where significant emphasis has been placed on using technology as a key weapon in the fight against disease outbreaks and the promotion of child health. In particular, mobile technologies are viewed by the Ministry of Health and Sanitation (MoHS) in Sierra Leone as an integral component of overall public health strategy with many mHealth initiatives being launched over the past few years by both the MoHS and a variety of non-governmental organizations (NGOs). One example is the use of MOTECH mobile health application, which is a result of collaboration between World Vision Ireland, World Vision Sierra Leone, Dimagi and Grameen Foundation. These mHealth systems are primarily designed to be used by community health workers (CHWs) who are typically the backbone of healthcare systems in the Global South (Babughirana et al., 2018). MOTECH is a mobile application developed to be used by CHWs to improve maternal, newborn and child health across the Bonthe District of Sierra Leone. This system implements the Timed and Targeted Counselling (TTL) initiative to improve outcomes for maternal and child health and nutrition.

Another example from Sierra Leone is the development of the Mobile Training and Support service (MOTS) to deliver refresher training to CHWs on the topic of vaccines and outbreak response. MOTS is an open-source platform which provides refresher training via an interactive voice response (IVR) system in the participant's preferred language (Babughirana et al., 2018). It provides mobile training to CHWs via their mobile phones as the basic requirement (MoHS, 2017). MOTS was initially developed under the Ebola vaccine deployment acceptance and compliance (EBODAC) programme and piloted with CHWs to promote the acceptance and uptake of Ebola vaccines (Mc Kenna et al., 2019; Babughirana et al., 2018). However, the emergence of COVID-19 has introduced another relevant application of MOTS due to the critical need to inform CHWs of health and medical protocols related to the virus without bringing them together for in-person training.

## 4.    METHODOLOGY

This paper adopts a qualitative methodological and interpretivist philosophical approach to address the research question. Initially, we rely on 5 semi-structured interviews which have been designed to ascertain how designers and developers of the mHealth initiative in Sierra Leone perceive the role of socio-technical factors as a means to improve user engagement. The COM-B model (Figure 1) is then used to map the facilitators and barriers of mHealth application engagement to understand users' behaviour of engaging with the mHealth app, with this being discussed further in section 4.2 below.

### 4.1 Data Collection and Methodology

A total of 5 semi-structured interviews were conducted during February and March 2021 with people who have been involved with the mHealth project in Sierra Leone. These included mHealth project managers, mHealth development facilitators, and others with a detailed knowledge of the project. The interviews were digitally audio-recorded, transcribed, and thematically analyzed. Thematic Analysis is a method for examining and classifying patterns of meaning in a dataset (Braun & Clarke, 2006). This involved the use of codes and themes to interpret and describe phenomenon under study. The codes are created, and themes are generated from collating codes to get acquainted with data (Ghasemi and Rasekh, 2020). Codes refer to an idea or feeling expressed in that part of a phrase in the data with the aim of organising data for subsequent interpretation, while themes refer to a specific ideas and patterns of meaning that come up repeatedly from the codes, that captures something significant about the data or/and research questions (Maguire and Delahunt, 2017). According to Braun and Clarke (2006: p. 78), thematic analysis "*provides a flexible and useful tool, which can potentially provide a rich and detailed, yet complex account of data*". The analysis of the interview data is guided by the six-step model of thematic analysis outlined by Braun and Clarke (2006).

According to Maguire and Delahunt (2017, p. 3352), "*thematic analysis is the process of identifying patterns or themes within qualitative data*". Since the six steps model of the thematic analysis described refers to themes, the notion of how the twelve themes were generated is explained in detail. The first step was to have a thorough overview of all the interview data about identifying what drives users to engage/disengage in the use/disuse of mHealth applications in order to design a better design process to improve user engagement in mHealth technologies. Hence, the concern is to address and analyse the data with the research question in mind. Initial codes were generated by highlighting phrases or sentences in the interview and shorthand labels (codes) were devised to describe their content. Coding reduces the data into small chunks of meaning. The codes were developed and modified through the coding process without using pre-set codes. Coding was done using NVIVO 12 (details in Appendix A), and the intention is not to code the content of the entire data set. The interview transcripts were coded separately by two persons. Each of the codes was compared, discussed, modified before moving to the rest of the transcripts. Every piece of the text was not coded; however, each coded transcript was relevant to or specifically addressed the research question.

The next step was to examine the codes created and search for a theme – an idea or concept that captured and summarised the meaningful pattern and recurring pattern in the data. Braun and Clarke (2006) clarified that there are no certain rules about what makes a theme. A theme is described by its significance. At this stage, codes were examined to ensure that they fitted together into a broader theme in order to address the research question. The next step was to name, review and refine the themes. Data associated with each theme were read to consider whether the data really did support the theme and how the themes work both within a single interview and across all the interviews. Naming themes involved providing a concise and easily understandable name for each theme. Themes were thus extracted from the transcripts of the interviews until we concluded that more themes could not be extracted from the data.

These interviews aimed at identifying what drives users to engage/disengage in the use/disuse of mHealth applications in order to design a better design process to improve user engagement in mHealth technologies. This obviously determined the interview questions and the analysis of the data. The interview participants were clear and consistent about what facilitates users' engagement with mHealth technologies in the transcripts. These facilitators generated the following themes: involving user early in the design stages of the mHealth, providing social connectedness; offline functionality; practice test; user manual or guidance and statistical information; and, tangible and intangible rewards. Each of these themes was classified as either socio-technical or technical in order to determine whether the socio-technical factors were deemed important for user engagement with mHealth in the transcript (as shown in Figure 2).

In identifying facilitators to user engagement, participants also highlighted barriers to user engagement. These were grouped into the following themes: cognitive overload; non-user-friendly design; cultural dimensions; lack of encouragement; perceived non-utility; and lack of app skills. In the final step, each of these themes is classified as well, as either socio-technical or technical in order to determine whether the socio-technical factors were deemed important for user engagement with mHealth in this case (details in Appendix B).

## 4.2 COM-B Model

The themes identified were analyzed with the Capability, Opportunity, Motivation – Behaviour (COM-B) model (Michie et al., 2014) to understand users' behaviour of engaging with the mHealth app. By considering user engagement with an app as a behaviour, the COM-B model provides a broad framework for understanding mHealth application engagement. According to the COM-B model, behaviour (e.g. user engagement with app) arises from the interaction between the individual's capability, both physical (e.g. app skills) and psychological (e.g. knowledge of using an application), their opportunity to behave in a certain way, both physical (e.g. via features of the app) and social (e.g. resulting from recommendations to use an app), and their motivation to behave, both automatic (e.g. emotional rewards from using an app) and reflective (e.g. belief in the benefits of the app). Further details of the mapping process and the elements of the COM-B model are discussed in more detail in the following section.

## 5.     FINDINGS AND DISCUSSIONS

The identified facilitators and barriers to mHealth application engagement in this case as derived from the interview data and mapped onto the components of the COM-B model are presented in Figure1.

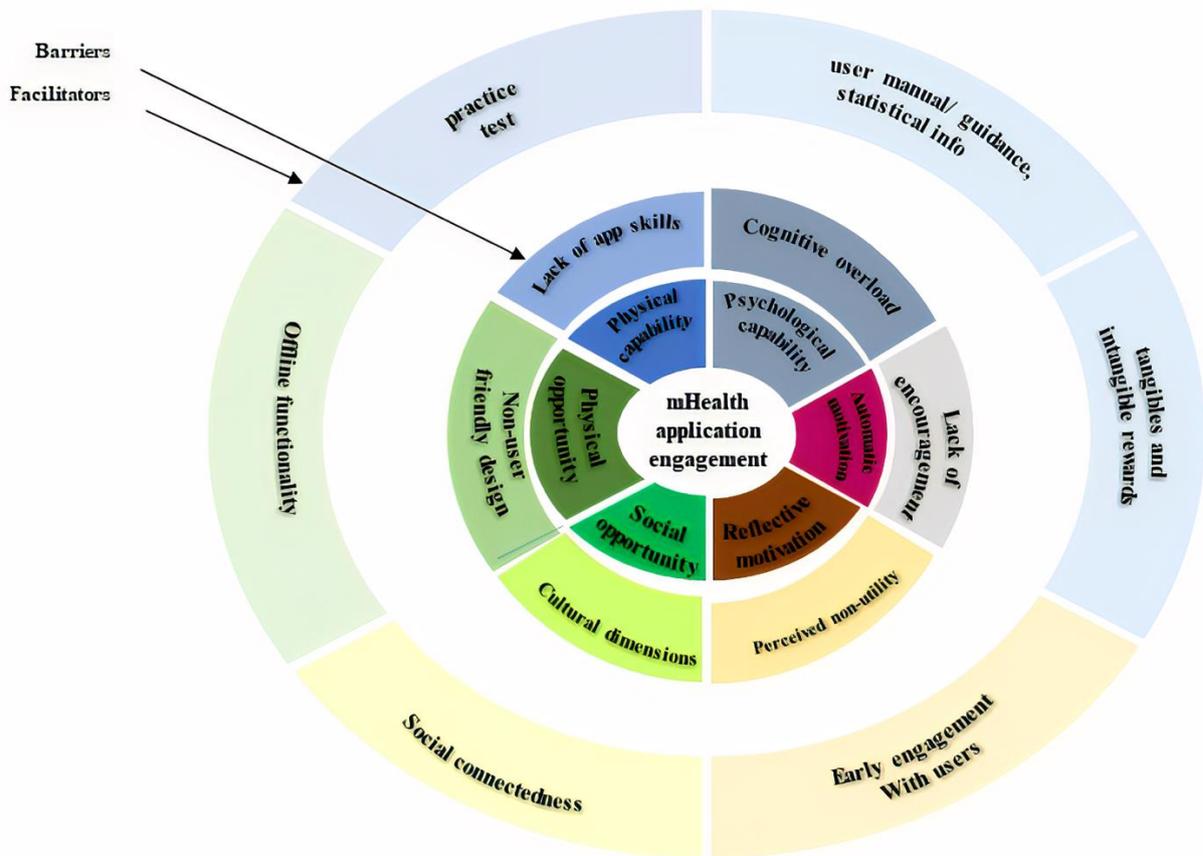

*Figure 1: A visual representation of how the facilitators and barriers of mHealth application engagement in the Sierra Leone mHealth case mapped onto the components of the COM-B model (Adapted from McDonagh et al. 2018, P. 8).*

The individual elements of the COM-B model and the outcome of the mapping process are now discussed in greater detail.

1. Physical capability refers to the appropriate skills and stamina that are required of users in order to take part in a particular behaviour. Many of the CHWs in this case struggle to engage with mHealth apps due to a lack of technical and language skills and low digital literacy. Several interviewees explained that CHWs expressed that the level of training provided to acquire technology skills shaped their level of commitment to engage with mHealth. This was highlighted by one interviewee as follows:

*"Ministry of Health and Sanitation in Sierra Leone provided training for CHWs to acquire skills to engage with mHealth apps in language of prefers, which helped to enhance learning uptake and engagement" (D2).*

This indicates that CHWs could be helped to more fully engage with the mHealth app by offering app-use tutoring.

2. Psychological capability refers to a user's knowledge that is required to engage in a particular behaviour. All interviewees identified interface minimalism—i.e making app interface as simple as it needs to be, as a factor that facilitate user engagement. A fewer elements on an interface results in lower cognitive load for users This factor is consistent with previous literature that reported engagement with health apps is impacted by factors affecting users' capabilities that include different types of knowledge such as user guidance, statistical information, health information (Szinay et al., 2020; Baumeister et al., 2014) and reduced cognitive load (Szinay et al., 2020;

Lyzwinski et al., 2018; Szinay et al., 2021). The education level of users was highlighted by one respondent:

*"When the MOTS mHealth solution was to be designed, the idea was to use smart phone but because of the level of education of users, basic features phones were recommended so that the CHWs can easily relate with" (D3).*

Furthermore, some of the interviewees reported that clear instructions on how to increase capability to perform a behaviour such as accessing the MOTS system by CHWs affects their engagement with an app. This is evidenced by the following quote:

*"When a CHW delays in responding to instruction given by the system when accessed due to lack of clear instructions, call ends if phone key is not immediately pressed/navigated after instruction" (D3).*

This means that these CHWs could be provided with clear instructions on what they need to do to achieve a given task, to get fully engaged with the mHealth app.

3. Physical opportunity refers to the set of circumstances that make it easier for the users to engage with a behaviour. All the interviewees highlighted that an opportunity for two-way communication between the mHealth app and user, as well as user-friendly design and interaction, facilitates user engagement in this case. This aligns with the previous literature that explains how apps can be improved by targeting the design and engagement features, such as user-friendly design, or health professional support (Zhao et al., 2016; Coughlin et al., 2016).

Some of interviewees suggested that careful selection of the terminology used to explain the app and what it does, such as simple and clear local language, and pictorials, creates an impact on user engagement.

*"The terminology used within the app, such pictorials, language and aesthetic can affect user acceptability and engagement" (D2).*

In addition, the need for offline functionality was identified as a physical opportunity factor impacting user engagement. This was highlighted by the example provided by one of the interviewees about accessing the MOTS system from a weak or poor network connection, and how this negatively affects CHWs from engaging with it. In addition, an interviewee identified free network access and working with data in offline mode even when CHWs do not have access to internet as an important factor for engagement:

*"CHWs can use MOTS system without paying for it and work offline mode using Bluetooth technology to upload information to DHIS system" (D1).*

This indicates that these CHWs could be assisted to engage with the mHealth app more fully by providing free internet and Bluetooth technology where there is no or poor network connection.

4. Social opportunity refers to a user's social circle enabling and facilitating a behaviour. Users' sharing of knowledge and experiences of their mHealth app engagement within their social circle makes it easier for them to engage with mHealth app. The possibility for community health workers to share knowledge and experiences within mHealth apps was considered important social support that facilitates engagement with mHealth apps. According to Puszkiewicz et al., (2016), this type of social circle was found to improve intention to engage with a mobile app intervention designed for regular participation for physical activity during and after cancer treatment.

In order to ensure that there is social opportunity, we need to study users' culture and capture it into mHealth apps. Here, culture is defined according to Ford and Kotzé (2005, p.716) as *"the patterns of thinking, feeling and acting that influence the way in which people communicate amongst themselves and with computers".* Culture is divided into two layers – objective and subjective (Stewart & Bennett, 2011). Objective culture means that intended meaning of user interface representations, such as symbols, icons and language, are translated to suit the target cultures, so that they are understood correctly (Ford and Kotzé, 2005). While subjective culture ensures that interface representations reflect the values, ethics and morals of the target users (Russo and Boor, 1993). This is evidenced by one particular comment from the interview as follows:

*"Social and Cultural dimension needs to be incorporated in considerations of user engagement designs for user acceptability before deployment" (D4).*

This indicates that the user's social circle, or community of practice, who also use the mHealth app could assist with engagement. The social circle should be considered by integrating the social norms and cultures of users at the design and development stages of the mHealth app.

5. Automatic motivation refers to the user's reinforcement and emotions that sustain engagement with the mHealth app. All the interviewees stated that offering rewards and various non-financial incentives was found to be a useful way to increase engagement. This is also consistent with the literature (Anderson et al., 2016; Perski et al., 2017; Perski et al., 2018), and this type of motivation is consistent with previous literature that observed that users found intangible rewards (e.g. badges) motivating (Peng et al, 2016), while others would want to receive tangible rewards instead (e.g. gift cards, cash, reduction in health insurance or vouchers provided by hospitals) (Peng et al., 2016; Baskerville et al., 2016). Our interviewees confirmed this by stating the following about bonuses given to CHWs:

*"At the end of the year or month CHWs that have performed very well are given bonus which could be in a form of token payment, boots or any other programmatic tools as a way of acknowledging them as outstanding users" (D4).*

This means that these CHWs could be provided with reinforcement and encouragement that will invoke their positive emotions to get them fully engage with the mHealth app.

6. Reflective motivation refers to user's beliefs that their needs and values are reflected in the mHealth apps. Some of the interviewees observed that lack of perceived utility of the app during the design can hinder user engagement. Perceived utility refers to where there is no disparity between what the users' needs and what an app offers. Siznay et al. (2019, p.39), observed that "*unmet expectations of an app would lead to disengagement and frustration with the app*". This was particularly apparent in one interviewee's comment on CHW engagement with mHealth indicators*:*

*"Users of mHealth apps such as CHWs may have issues engaging with mHealth apps if they are not involved in making sure that the indicators of their needs are reflected in the mobile apps, involving users to reflects their needs at the design stages of the mHealth apps, boosts their confidence of using mHealth apps" (D2).*

All the facilitators and barriers of engagement with mHealth interventions that were identified from from the thematic analysis discussed above were classified as either technical or socio-technical factors as shown in Figure 2. This was achieved using thematic analysis (as specified in section 4.1) to identify the technical and socio-technical facilitators and barriers to user engagement with mHealth in this case.

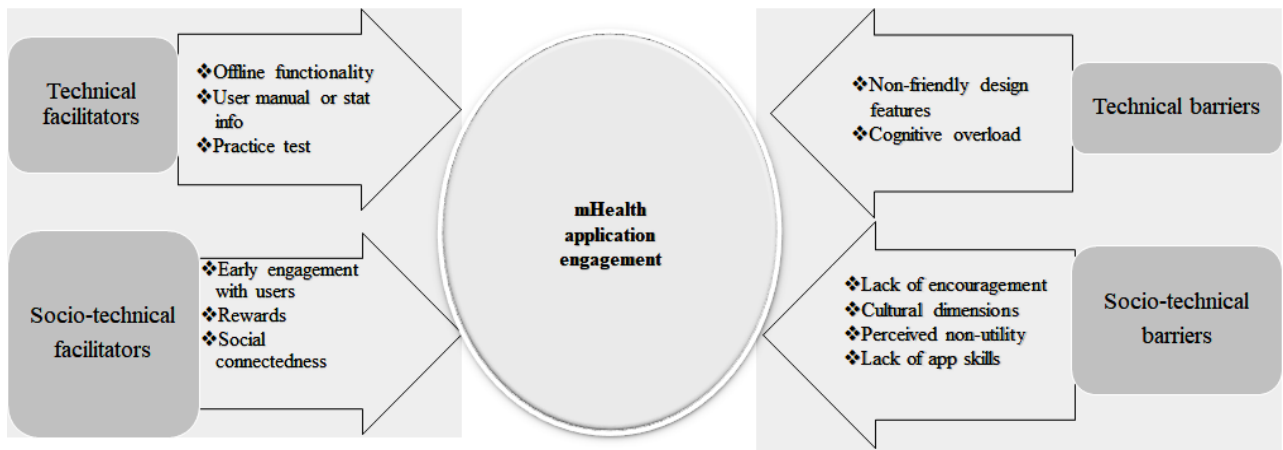

**Figure 2**: *Classifying facilitators and barriers identified from the key findings into socio-technical and technical factors.*

The presence of these technical and socio-technical facilitators and barriers of user engagement with mHealth apps are important because they hold key position for mobile health system to sustain and scale as indicated by the COM-B analyses of data collected from interviewees. This aligns with the previous literature that discusses mHealth sustainability and scaling (Gagnon et al., 2016; Putzer & Park, 2012; Koivunen & Saranto, 2018). Therefore, the presence of only the techno-centric facilitators and barriers of user engagement with mHealth apps cannot offer an improved user engagement.

The facilitators and barriers identified, and the result of the COM-B analysis completed, has informed the development a new mHealth design framework called **De**sign Pro**c**ess **E**ngagement E**n**hancement Sys**t**em (DECENT). DECENT will allow designers/developers uncover user engagement impacting factors in order to improve user engagement. DECENT is discussed in detail in the following section.

# 6.    PROPOSED DESIGN APPROACH FOR USER ENGAGEMENT - DECENT

DECENT will help inform mHealth designers and developers to understudy mHealth context and uncover user needs to improves user engagement with mHealth interventions. DECENT will serve as a guide towards making decisions in the design, development, implementation of mHealth for user engagement in the Global South.

DECENT uses design science research and use of theories to uncover socio-cultural contexts of the user group as shown in Figure 3. The design science research is guided by the three research cycles outlined by Hevner (2007). The first research cycle is the relevance cycle, where the environment of the end-user is understood. The second cycle is design cycle, where artefacts that are relating to solution of research problem are created and assessed. The third cycle is rigour cycle, where findings from evaluating proposed solutions form part of the existing knowledge base.

It is established in this study that technical and socio-technical facilitators and barriers of user engagement with mHealth apps are important because they hold key position for mobile health system. Our future work will present a new framework – DECENT – that will help mHealth designers and developers discover the key factors impacting user engagement with mHealth apps and will serve as a guide towards making decisions in the design, development, implementation of mHealth for user engagement in the Global South. The proposed structure of DECENT is shown in Figure 3.

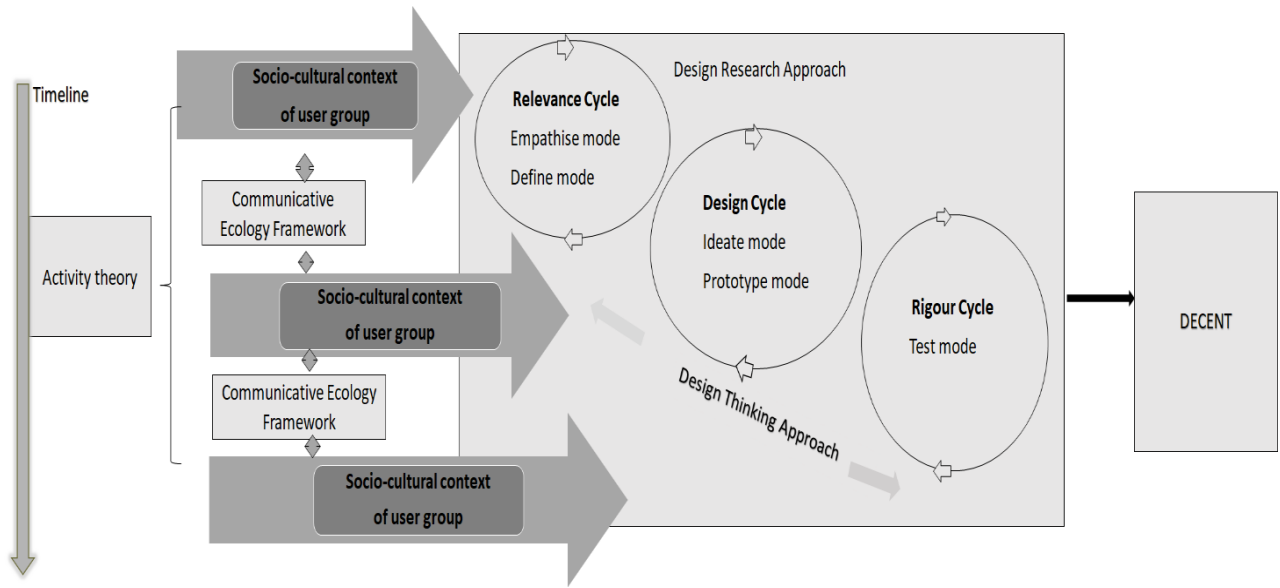

***Figure 3:** The **De**sign Pro**c**ess **E**ngagement **En**hancement Sys**t**em (DECENT) model.*

DECENT will incorporate various theories which will help designers and developers to uncover and take account of socio-technical factors that impact user engagement with mHealth. One of the theories we propose to use is activity theory. Activity theory is a tool which can be used to uncover the socio-technical context of mHealth implementations. We propose to leverage activity theory to support the creative design process and to provide a theoretical framework of social and cultural contexts in the design process of DECENT for user engagement. Activity theory is embodied with primary units that aims towards incorporating social-technical content for user centered design.

Activity theory presents a robust framework for studying contextual factors, and it shows us the complexities and fluidity of activities in context. However, it does not tell us how activities are structured by contextual factors. The vision of context and culture here is still limited. For example, using activity theory does not tell us how user engagement with mHealth technologies are structured in social contexts. Thus, the Communicative ecology framework (CEF) will be applied (as described in Foth & Hearn, 2007) which integrates three layers of interpretation (technical, social and discursive) to deliver a rich description of how mHealth is structured in a social context.

## 7.    CONCLUSIONS

This work contributes to the advancement of knowledge in the mHealth field by showing that socio-technical aspects of mHealth design are important for user engagement. However, it is not enough to simply analyse mHealth design from a socio-technical perspective. Thus, this research proposes a new DECENT framework for designing mHealth for user engagement which will be developed in order to aid inclusion of socio-technical factors in mHealth. This will allow users' needs and goals to be taken into account and advance identification of problems concerning lack of user engagement.

Although this research has provided promising results, part of our future work will discuss our proposed DECENT framework with a wide variety of stakeholders following the steps highlighted in this work. In addition, we will make a particular effort to engage as fully as possible with the CHWs in Sierra Leone as part of this process. Finally, we call on the wider IFIP WG 9.4 community to engage more fully with the concept of user engagement in information systems and mHealth in the Global South.

## ACKNOWLEDGEMENTS

This research was conducted with the financial support of Science Foundation Ireland under Grant Agreement No. grant 18/CRT/6222 at the ADAPT SFI Research Centre at Trinity College Dublin. The ADAPT SFI Centre for Digital Content Technology is funded by Science Foundation Ireland through the SFI Research Centres Programme and is co-funded under the European Regional Development Fund (ERDF) through Grant #13/RC/2106_P2

# Appendix A: Coding

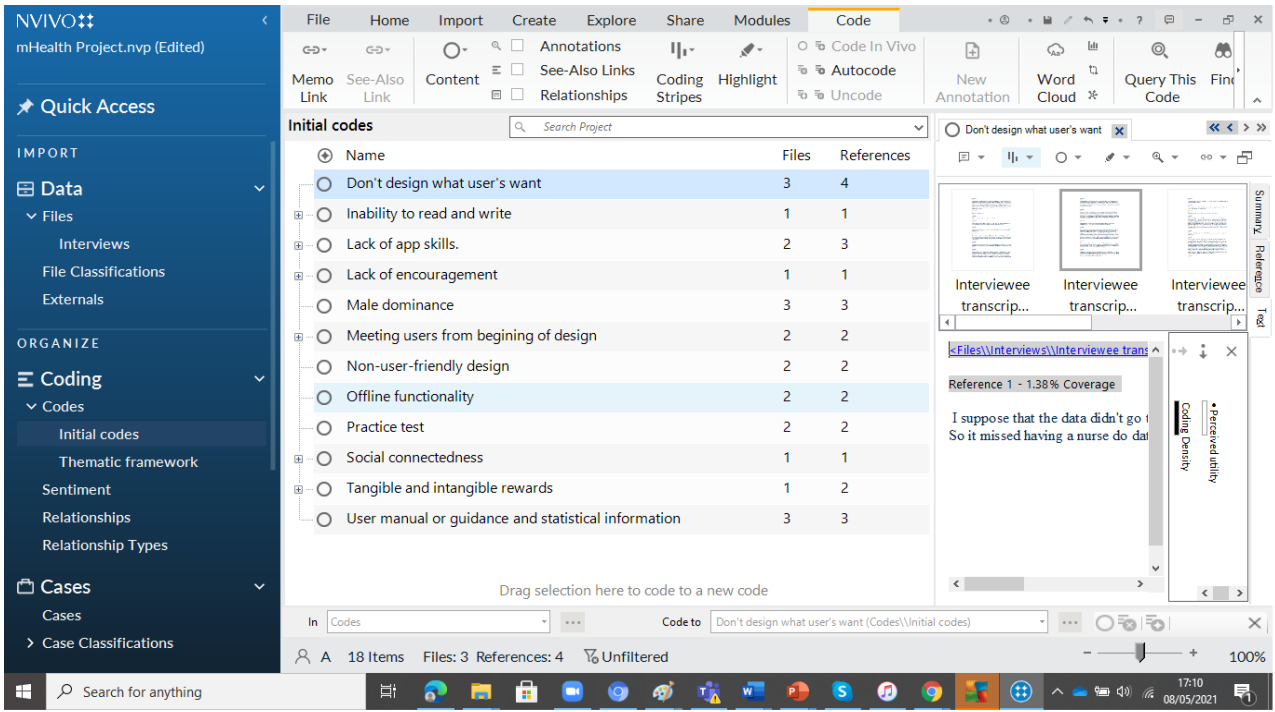

# Appendix B: Extracting themes from codes

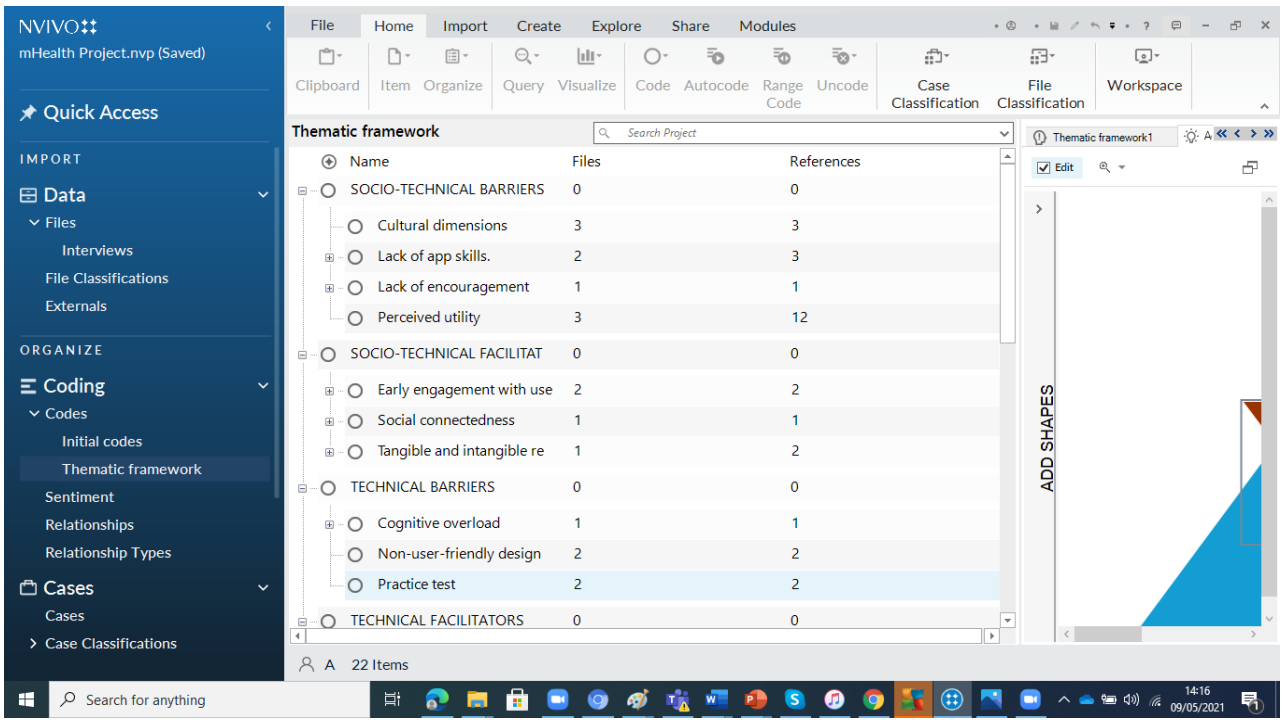